\documentclass[paper, traditabstract]{aa}
\usepackage{setspace}
\usepackage{txfonts} 
\usepackage{textcomp}
\usepackage{natbib}
\usepackage{amssymb}
\usepackage{graphicx}
\usepackage{tabularx}
\usepackage[T1]{fontenc}
\usepackage{multirow}

\def\HI{$\ion{H}{I}$}
\def\halpha{\ion{H}{$\alpha$}}

\newcommand{\kms}{$\,$km$\,$s$^{-1}$}
\newcommand{\ergs}{$\,$erg$\,$s$^{-1}$}
\newcommand{\ergscm}{$\,$erg$\,$s$^{-1}\,$cm$^{-2}$}

\newcommand{\mJybeam}{mJy beam$^{-1}$}
\newcommand{\msun}{{${\rm M}_\odot$}}
\newcommand{\msunyr}{{${\rm M}_\odot$ yr$^{-1}$}}

\def\OIII{[O{\,\small III}]}

\def\emph#1{{\sl #1}}
\newcommand{\ltsima} {$\; \buildrel < \over \sim \;$}
\newcommand{\gtsima} {$\; \buildrel > \over \sim \;$}
\newcommand{\lta} {\lower.5ex\hbox{\ltsima}}
\newcommand{\gta} {\lower.5ex\hbox{\gtsima}}

\begin{document}

   \title{What triggers a radio AGN?}

   \subtitle{The intriguing case of \mbox{PKS B1718--649}}

   \author{F. M. Maccagni
          \inst{1}$^,$\inst{2}
          \and
          R. Morganti\inst{1}$^,$\inst{2}
          \and
          T. A. Oosterloo\inst{1}$^,$\inst{2}
          %
          %\thanks{Just to show the usage
         % of the elements in the author field}
         \and
         E. K. Mahony\inst{2}
          }

   \institute{Kapteyn Astronomical Institute, Rijksuniversiteit Groningen, Landleven 12, 9747 AD Gronignen, The Netherlands\\
              \email{maccagni@astro.rug.nl}
         \and
             Netherlands Institute for Radio Astronomy, Postbus 2, 7990 AA, Dwingeloo, The Netherlands\\
             %\email{[morganti;oosterloo]@astron.nl}
            % \thanks{The university of heaven temporarily does not
                     %accept e-mails}
             }

%   \date{Received September 15, 1996; accepted March 16, 1997}

% \abstract{}{}{}{}{} 
% 5 {} token are mandatory
 
  \abstract
   {
We present new Australia Telescope Compact Array (ATCA) observations of the young ($<10^2$ years) radio galaxy \mbox{PKS B1718--649}. We study the morphology and the kinematics of the neutral hydrogen ($\ion{H}{I}$) disk ($M_{\rm{\ion{H}{I}}}=1.1\times10^{10}$ M$_\odot$, radius$\sim30$ kpc). In particular, we focus on the analysis of the cold gas in relation to the triggering of the nuclear activity. The asymmetries at the edges of the disk date the last interaction with a companion to more than $1$ Gyr ago. The tilted-ring model of the $\ion{H}{I}$ disk shows that this event may have formed the disk as we see it now, but that it may have not been responsible for triggering the AGN. The long timescales of the interaction are incompatible with the short ones of the radio activity. In absorption, we identify two clouds with radial motions which may represent a population that could be involved in the triggering of the radio activity. We argue that \mbox{PKS B1718--649} may belong to a family of young low-excitation radio AGN where, rather than through a gas rich merger, the active nuclei (AGN) are triggered by local mechanisms such as accretion of small gas clouds. }

   \maketitle
%
%________________________________________________________________

\section{Introduction}

Active galactic nuclei (AGN) are associated with the accretion of material onto a supermassive black hole (SMBH). There are tight scaling relations between the mass of the SMBH and the stellar bulge mass of the host galaxy, the stellar velocity dispersion and the concentration index (see~\citealt{magorrian,ferrarese,hopkins}). This suggests that the energy emitted by the AGN strongly affects the evolution of its host. Conversely, the phenomena going on within the host galaxy may set the conditions to trigger the AGN. It still remains very unclear, though, what are these triggering mechanisms and when they occur. Gas and dust are the fuel of the nuclear activity and are now being detected in many elliptical galaxies (\citealt{oosterloo,serra}), i.e. the typical host of radio AGN . Hence, insights on the triggering mechanisms of AGN can perhaps be found observing the cold gas in its host: through the fuel we understand how the AGN was started. 

Radio AGN can be classified in two distinct populations based on the intensity ratios of their optical emission lines: low-excitation radio galaxies (LERG) and high-excitation radio galaxies (HERG). This dichotomy is reflected in the different efficiency of the accretion into the black hole: LERG are radiatively inefficient accretors, with most of the energy from the accretion being channeled into the radio jet, while HERG are radiatively efficient (\citealt{best}), showing strong evidence of nuclear activity also in the optical band. It is possible that these two accretion modes are triggered by gas in different initial physical conditions (\citealt{hardcastle,kauffmann2009,cattaneo2009}). In merger and interaction events the cold gas of the galaxies may loose angular momentum and form a radiatively efficient accretion disk around the black hole \citep{smith,sabater}. Internal slow processes, known as `secular' (\citealt{kormendy}), are also able to drive cold gas into an accretion disk. 

On the other hand, radiatively inefficient accretion may occur through, e.g. accretion of hot coronal gas  \citep{allen,hardcastle,balmaverde}.
% via the Bondi mechanism. This is the most convenient analytical solution to determine the accretion rate into a BH in the inefficient mode, but it assumes the hot gas to be symmetric, adiabatic and steady, which almost never occurs in galaxies. 
Nowadays simulations \citep{soker,gaspari2012,gaspari}, take into account more physical conditions, i.e. radiative cooling, turbulence of the gas, and heating, and show that cold gas clouds may form through condensation into the hot halo of the host galaxy. Through collisions the clouds may loose angular momentum and chaotically accrete into the black hole, possibly triggering a radiatively inefficient AGN.

Neutral hydrogen (\HI) observations of radio galaxies are
particularly suitable for understanding the connection between the fueling
and the triggering of the AGN. The kinematics of the (\HI) can trace if and when a galaxy has undergone an interaction event, as well as the presence of ongoing secular processes and in-falling clouds. Thus, by comparing the timescales of these events with the age of the radio source, it is possible to learn about the mechanism responsible for the triggering of the AGN. In some of these objects, see e.g. \citealt{emonts1,struve,shulevski}, large delays have been found between these events, suggesting that the recent radio-loud phase has been triggered by the chaotic accretion of cold clouds.\\
%Such AGN may reach accretion rates of $\sim0.1$~\msunyr, on short timescales $\sim10^6$~Myr. , typically, of low luminosity.
%it seems most likely that the LERG population simply does not have a radiatively efficient accretion flow, and so produces none of the optical or X-ray characteristics of a conventional AGN. LERG may be a class of luminous active galaxies that accrete radiatively inefficiently, with almost all the available energy from accretion being channeled into the jets.
%These processes are the combination of tidal interactions (\citealt{kuo,ellison}), bars (\citealt{ho,lee}), and nuclear spirals (e.g.,~\citealt{norman,macio,storchi}).
\indent Absorption studies of the neutral hydrogen are successful in tracing the cold gas in the inner regions of the AGN. There, the dynamical time of the gas is similar to the lifetime of the radio source (see e.g.~\citealt{heckman,shostak,vg,morganti2001,vermeulen,gupta}). These studies show that the $\ion{H}{I}$, in many AGN where it is detected, is distributed in a circum-nuclear disk/torus around the active nucleus (\citealt{gereb}, and references therein). In some radio galaxies, clouds of $\ion{H}{I}$ not following the regular rotation of the disk are detected. These clouds can be outflowing from the nucleus e.g. pushed by the radio jet (e.g. \mbox{4C 12.50} and  \mbox{3C 293} (~\citealt{morganti2005,morganti2013,mahony}) or  falling into the nucleus, possibly providing fuel for the AGN (see the case of NGC~315, \cite{morganti2009}, and 3C~236, \cite{struve}).  Thus, the \HI~ can trace different phenomena in the nuclear regions.

Young radio sources ($<10^5$ years) are ideal to study the cold gas in AGN, and, in particular, its role in their triggering. Among all kinds of radio galaxies, the $\ion{H}{I}$ is detected more frequently in these sources (\citealt{emonts,odea,gereb}), suggesting that the cold neutral gas must play an important role in the early life of a radio AGN. 

In this paper, we present the Australia Telescope Compact Array (ATCA) observations of a young $\ion{H}{I}$ rich radio source: \mbox{PKS B1718--649}. We analyze the kinematics of the neutral hydrogen, detected in absorption and in emission, to understand what may have triggered the radio activity.

The paper is structured as follows. In Sect.~\ref{sec:galaxy}, we describe the overall properties of \mbox{PKS B1718--649}. Our observations and data reduction are described in Sect.~\ref{sec:datareduction}. In Sect.~\ref{sec:dataanalysis} we describe the main properties of the $\ion{H}{I}$ disk retrieved from our observations. In Sect.~\ref{sec:discussion} we illustrate the tilted-ring model of the $\ion{H}{I}$ disk retrieved from our data and we compare the timescales of the merger to the ones of the radio activity. Section~\ref{sec:abs} focuses on the analysis of the $\ion{H}{I}$ detected in absorption, which could be connected to a cloud close to the center of the galaxy, possibly interacting with the nuclear activity. In the last Section, we summarize our results and provide insights on the accretion mode of this radio source. 

Throughout this paper we use a $\Lambda CDM$ cosmology, with Hubble constant $H_0= 70$ km s$^{-1}$/Mpc and $\Omega_\Lambda=0.7$ and $\Omega_M=0.3$. At the distance of \mbox{PKS B1718--649} this results in 1 arcsec $= 0.294$ kpc.

%which both are powerful ($P_{\rm{1.4GHz}} >10^{25}$W/Hz) and compact ($R<1$ kpc) sources
%__________________________________________________________________

%, lead NGC6328 to be classified as S0 (\citealt{lauberts}) or as SABab with luminosity class III (\citealt{devauc})
%\citealt{savage}
%The distance from lobe to lobe is only $2$ pc with components of at PA $135^\circ$.
%a total $H_2$ mass of $\sim0.12\times10^7$ M$_\odot$ and

\section{Properties of \mbox{PKS B1718--649}}
\label{sec:galaxy}

\begin{figure}
   \centering
  \includegraphics[width=0.4\textwidth]{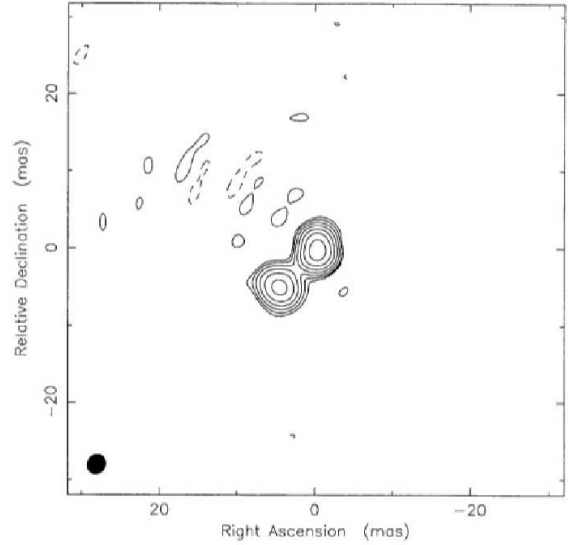}
   \caption{4.8 GHz SHEVE VLBI image of \mbox{PKS B1718--649}. The contour levels shown are -1,1,2,4,8,16,32 and $64\%$ of the peak flux density of 2.0 Jy beam$^{-1}$. The restoring beam FWHM dimensions are 2.6x2.3 mas, the major axis position angle is $-29.5^\circ$ [\citealt{tingay}].}
     \label{fig:vlbi}
   \end{figure}

\begin{figure}
   \centering
  \includegraphics[width=0.48\textwidth]{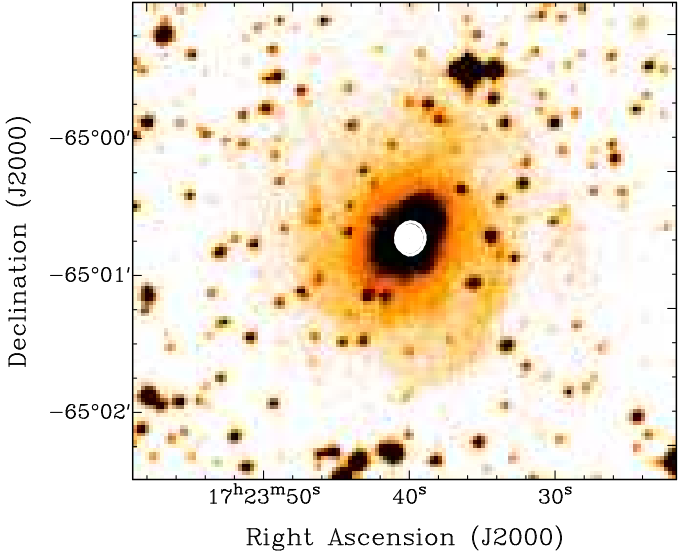}
   \caption{I-band Image of \mbox{PKS B1718--649}: the spiral arms form an envelope around the elliptical like central part of the galaxy [UKSchmidt telescope, DSS, 1993 (NED)]. The unresolved radio source is marked in \emph{white}.}
     \label{fig:opt_1}
   \end{figure}

\begin{figure}
   \centering
  \includegraphics[width=0.48\textwidth]{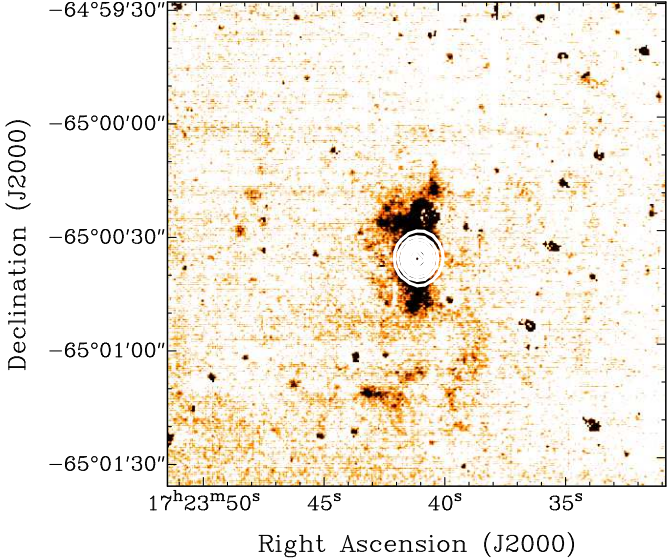}
   \caption{H$_\alpha$ image of \mbox{PKS B1718--649} from the CTIO 1.5 m telescope. The field is 2.3'$\times$2.3' with North at the top. A complex extended emission region is visible, with a strong concentration N-S from the nucleus. An extensive filamentary structure seems to follow the spiral structure seen in the I band [\citealt{keel}]. The unresolved radio source is marked in \emph{white}.}
     \label{fig:opt_2}
   \end{figure}

\begin{figure}
   \centering
  \includegraphics[trim = 30 0 30 0, clip,width=0.48\textwidth]{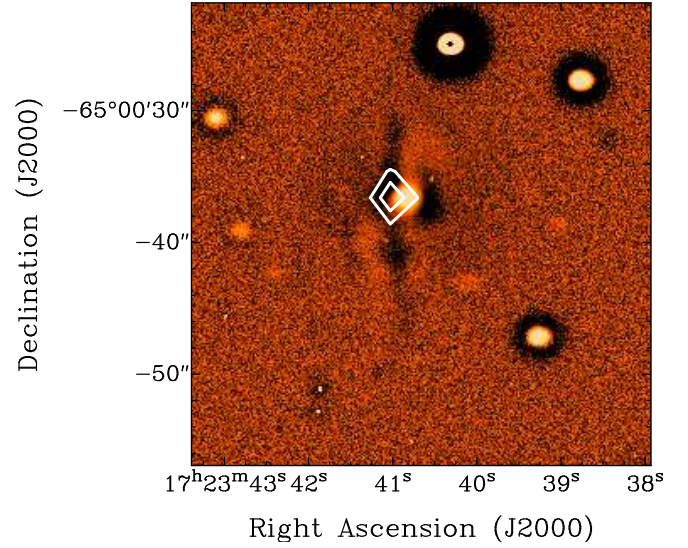}
   \caption{Detail of the GUNN-I observation of \mbox{PKS B1718--649}. It is strongly suggested the presence of an absorbed thin disk oriented N-S along the galaxy bulge. We applied sharp masking to the image to better highlight the dust lane [\citealt{veron}]. The unresolved radio source is marked in \emph{white}.}
     \label{fig:opt_3}
   \end{figure}

The radio source \mbox{PKS B1718--649} is a young Giga-Hertz Peaked (GPS) radio source. GPS are defined as compact radio sources (smaller than their optical host), not core dominated, with a spectrum peaking in the GHz region (\citealt{fanti}). The compactness of the source and the peak of the spectrum indicate the very young age of these sources. In \mbox{PKS B1718--649}, VLBI observations at $4.8$ GHz (\citealt{tingay}) show a compact double structure (size $R\sim2$ pc; Fig.~\ref{fig:vlbi}(b)). Follow up observations (\citealt{giroletti}) measured the hot spot advance velocity and determined the kinematic age of the radio source to be $10^2$ years. \mbox{PKS B1718--649} is hosted by the galaxy \mbox{NGC 6328} (\citealt{savage}), whose physical parameters are given in Table~\ref{tab:parameters}. This galaxy has an early-type stellar population of stars and elliptical morphology. There is a faint spiral structure, visible in the I-band (Fig.~\ref{fig:opt_1}). The ionized gas (\halpha) follows this same spiral distribution in the external regions, but it is distributed along the N-S axis in the center of the galaxy (\citealt{keel}, Fig.~\ref{fig:opt_2}). The high-density \halpha~ in the inner regions is a sign of active star formation. \emph{Spitzer} Infra-Red data (\citealt{willett}) infer a star formation rate (SFR) of SFR$_{\rm Ne}=1.8\pm0.1$ \msunyr and SFR$_{\rm PAH}=0.8$ \msunyr. A dust lane is visible south of the nucleus oriented approximately at PA$=170^\circ$, (as shown by the ESO-SUSI optical observations, Fig.~\ref{fig:opt_3}). 
   
 \begin{table}[tbh]
 \renewcommand{\tablename}{Tab.}
\caption{Basic properties of \mbox{PKS B1718--649}.}
 %\caption[]{\label{nearbylistaa2}List of nearby SNe used in this work.}
\begin{tabular}{lcc}
 \hline \hline
 Parameter & Value & Ref.\\
  \hline
Morphological type & S0 / SABab III & (1,2)\\
M$_{K20}$ & $-24.4$ & (3) \\
Radio continuum center $\alpha$ (J2000) & $17^{\rm{h}}23^{\rm{m}}41.09^{\rm{s}}$ & (4) \\
Radio continuum center $\delta$ (J2000) & $-65^{\circ}00^{\rm '}37^{\rm ''}$ & (4)\\
Distance (Mpc)& $62.4$ ($z=0.0144$) & (5)\\ 
Size of the radio lobes (mas) &  $2.3\times1.2$ - $2.4\times1.5$ & (6) \\
Radio source PA ($^\circ$) & $135$ & (4) \\
Peak Flux [2.4 GHz] (Jy) & $4.4$ & (4) \\ 
Flux [1.4 GHz] (Jy) & $3.98$ & (*) \\
Radio Power [1.4 GHz] (W Hz$^{-1}$) & $1.8\times10^{24}$  & (*)\\
Age of the radio source (yr) & $10^2$ & (6) \\
Total $\ion{H}{I}$ Mass (M$_\odot$) & $1.1\times10^{10}$ & (*)\\
Radius $\ion{H}{I}$ disk (kpc) & $29$ & (*)  \\
$L_{\rm{H}_\alpha}$ (\ergs) & $2.9\times10^{41}$ & (7) \\
$M_{\rm{H}_2}$ (M$_\odot$)  & $0.2\times10^7$ & (8) \\
$M\star$ (M$_\odot$) & $4.9\times10^{11}$ & (*) \\
SFR$_{\rm PAH}$ (\msunyr) & $0.8$ & (8) \\
\hline
\end{tabular}
\tablebib{(1)~\citet{lauberts};
(2) \citet{devauc};  (4) 2MASS 2003;  (3) \citet{tingay}; (5) $\ion{H}{I}$PASS,~\citet{doyle};
(6) \citet{giroletti}; (7) \citet{keel}; (8) \citet{willett};
(*) this work.}
\label{tab:parameters}
\end{table}

%      \begin{figure*}
%   \centering
 %  \includegraphics[width=4.3cm]{../figures/galaxy.eps}
%      \includegraphics[width=4.3cm]{../figures/galaxyhalpha.eps}
%   \includegraphics[width=4.3cm]{../figures/galaxydustlane.eps}
%   \caption{\emph{(a)}:  I Image of \mbox{PKS B1718--649}: the spiral arms constitute an envelope around the elliptical like central part of the galaxy. [UKSchmidt telescope, DSS, 1993 (NED)] \emph{(b)}: H$_\alpha$ image of \mbox{PKS B1718--649} from the CTIO 1.5 m telescope. The field is 2.3'x2.3' with North at the top. A complex extended emission region is visible with a strong concentration N-S from the nucleus, as well as more extensive filamentary structure, that seems to be in agreement with the spiral arms structure seen in the I band. [\citealt{keel}. \emph{(c)}: Detail of the GUNN-I observation of PKSB1718-649 (\citealt{veron}). It is strongly suggested the presence of an absorbed thin disk oriented N-S along the galaxy bulge. [Veron-Cetty et al. 1997]. }
 %    \label{fig:opt}
  % \end{figure*} 
   %trim = 70 0 115 0, clip,trim = 60 0 110 0, clip,trim = 105 0 165 0, clip,

ATCA observations (\citealt{veron}) detected the massive ($M\gtrsim10^{9}$ M$_\odot$) $\ion{H}{I}$ disk of \mbox{PKS B1718--649}. As in $\sim20\%$ of early-type galaxies (\citealt{serra}), the disk appears fairly regular, except for slight asymmetries in the outer regions. The central depression, along with the orientation of the dust lane and the \halpha~region, are indicative of a warped structure of the disk. Two \HI~absorption lines were  detected against the radio core. The estimate of their optical depth was limited by the low velocity resolution of the observations.

   %The disk extends for R$\sim95$ arcsec, has an integrated flux $F_{\rm{1.4GHz}}\sim3.8$ Jy/beam, and mass .

%                                     Two column figure (place early!)
%______________________________________________ Gamma_1 (lg rho, lg e)
 %  \begin{figure*}
  % \centering
   %\includegraphics[width=3cm]{../figures/tingay_compact.eps}
   %%%\includegraphics{empty.eps}
   %%%\includegraphics{empty.eps}
   %\caption{a): 4.8 GHz SHEVE VLBI image of PKS1718-649. The contour levels shown are -1,1,2,4,8,16,32 and $64\%$ of the peak flux density of 2.0 Jy/beam. The restoring beam FWHM dimensions are 2.6x2.3 mas, the major axis position angle is $-29.5^\circ$. [Tingay et al. 1997].}
       %       \label{FigGam}%
    %\end{figure*}
%

%
%                                                One column figure
%----------------------------------------------------------- S_vib
%
%______________________________________________________________

%__________________________________________________ One column table
%

\section{Observations and data reduction}
\label{sec:datareduction}
The new $\ion{H}{I}$ observations of PKS B178-649 were taken with the ATCA in three separate runs of 12 hours. Full details are given in Table~\ref{tab:atca_obs}. The observations were centered at the $\ion{H}{I}$-line red-shifted frequency ($1.398$ GHz). For the three observations different array configurations were used in order to assure the best \emph{uv}-coverage. The Compact Array Broadband Backend (CABB) of the telescope provides a total bandwidth of $64$ MHz over $2048$ channels (dual-polarization). The primary calibrator for flux, phase and bandpass was PKS B1934-638 (unresolved and with a flux of $14.9$ Jy at $1.4$ GHz). A continuum image has been produced using the line-free channels.  At the resolution of our observations, the continuum is unresolved with a flux density of $S_{\rm{cont}}$=$3.98$ Jy. In order to track the variations in the bandpass and improve the calibration precision, we observed the calibrator for 15 minutes every 3 hours.

 \begin{table}[tbh]
\caption{Instrumental parameters of the ATCA observations.}
 %\caption[]{\label{nearbylistaa2}List of nearby SNe used in this work.}
\begin{tabular}{lc}
 \hline \hline
            Field Center (J2000) & $17^{\rm{h}}23^{\rm{m}}41.0^{\rm{s}}-65^{\circ}00^{\rm{m}}36.6^{\rm{s}}$   \\
            Date of the observations        & 11\rm{ Jun. }13; 04\rm{ Aug. }13; 10\rm{ Sep. }13 \\
            Total integration time (h) & 12 ; 12; 12     \\
            Antenna configuration & 6\rm{C}; 750\rm{D}; 1.5\rm{A} \\
            Bandwidth & 64~\rm{MHz};~2048~\rm{channels} \\
            Central Frequency &  1.398~\rm{GHz}\\
\hline
\end{tabular}

\label{tab:atca_obs}
\end{table}

The data were calibrated and cleaned using the $\tt{MIRIAD}$ package (\citealt{sault}). We joined together our 36 hours of observations with the 24 hours taken by\cite{veron}. For the $\ion{H}{I}$-line study, we fitted the continuum in the line-free channels with a $3$rd order polynomial. In \mbox{PKS B1718--649} the neutral gas is detected in emission and in absorption against the unresolved radio continuum.

The final data cubes to study the $\ion{H}{I}$ in emission and absorption have been produced using different parameters. For the study of the \HI~emission, to increase the signal-to-noise ratio (S/N), the final data cube considers visibilities only from 5 antennas, thus excluding the longest baselines. Robust weighting ($robust=0$) was applied to the data. The complete list of parameters used for the data reduction is presented in Table~\ref{tab:datareduction}. The restoring beam has a size of $29.7\times27.9$ arcsec and PA$ =16.6^\circ$. After Hanning smoothing, the velocity resolution is $\sim15$ km s$^{-1}$. The rms noise of the final data cube is $0.71$~\mJybeam. Hence, the minimum detectable ($3\sigma$) column density is $6.7\times10^{19}$ cm$^{-2}$, and the minimum mass is $3.8\times{10^7}$\msun .

To retrieve the best information from the $\ion{H}{I}$ detected in absorption, it is important to exploit the highest spatial and velocity resolution of the telescope. The final cube considers visibilities from all $6$ antenna. It has uniform weighting ($robust=-2$) and a restoring beam of  $11.21\times10.88$ arcsec. The velocity channels have a resolution of $6.7$ km s$^{-1}$. The noise in the cube is $0.94$~\mJybeam. The $3\sigma$ noise level gives the minimum detectable column density $9.5\times10^{17}$ cm$^{-2}$. In absorption we are more sensitive than in emission because of the higher velocity resolution of the data cube and because of the strong continuum flux of \mbox{PKS B1718--649}.

\begin{table*}
\caption{General data reduction parameters of the new ATCA observations of \mbox{PKS B1718--649}.}
\centering          
\begin{tabular}{l c c}     % 7 columns 
\hline\hline       
                      % To combine 4 columns into a single one 
Data cube parameters & Emission & Absorption\\\hline                    
Number of antennas & 5 & 6\\
Velocity resolution (km s$^{-1}$) & 15 & 6.7 \\
Angular resolution (arcsec) & $29.7\times27.9$ & $11.2\times10.9$\\
Beam PA (degrees) & 16.6 & 25.6\\ 
Weighting & 0 & -2(uniform)\\
Rms noise (mJy beam$^{-1}$) & 0.71 & 1.04 \\ 
Minimum detecable optical depth ($3\sigma$) & - & 0.0007 \\ 
Minimum detectable column density ($3\sigma$; cm$^{-2}$) & $6.73\times10^{19}$ & $9.5\times10^{17}$  \\
Minimum detectable mass ($3\sigma$; M$_\odot$ per resolution element) & $3.8\times10^7$& - \\
\hline                  
\end{tabular}

\label{tab:datareduction}
\end{table*}

\section{The neutral hydrogen in \mbox{PKS B1718--649}}
\label{sec:dataanalysis}
In this Section, we present the analysis of the resulting cubes from the combined datasets. The $\ion{H}{I}$ detected in emission reveals a regularly rotating disk. In absorption, we detect two separate lines. 

\subsection{$\ion{H}{I}$ emission}
\label{sec:emis}
\begin{figure}
   \centering
   \includegraphics[width=0.5\textwidth]{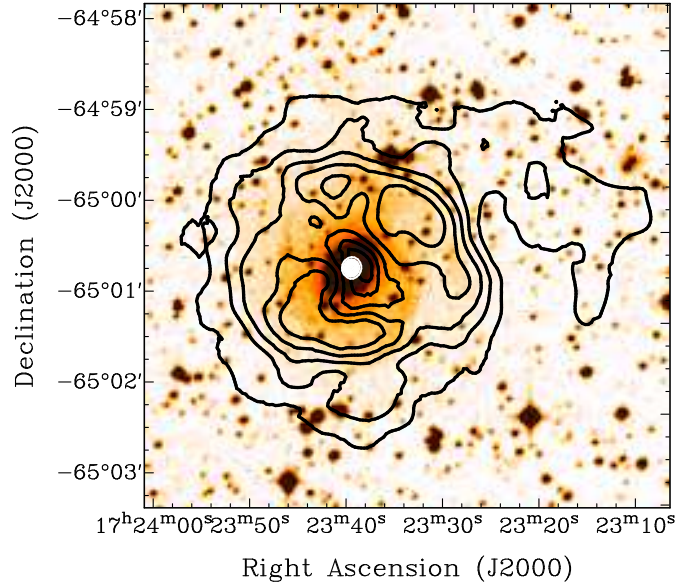}
   \caption{I-band optical image of \mbox{PKS B1718--649}, overlaid with the column density contours (in \emph{black}) of the neutral gas. The $\ion{H}{I}$ disk has the shape of an incomplete ring with asymmetries in the NW and in the S of the disk. The contour levels range between $7\times10^{19}$ cm$^{-2}$ and $8\times10^{20}$ cm$^{-2}$, with steps of $1.5\times10^{20}$ cm$^{-2}$. The unresolved radio source is marked in \emph{white}.}
     \label{fig:density} 
   \end{figure}

\mbox{PKS B1718--649} has a face-on $\ion{H}{I}$ disk extending beyond the stellar structure. Fig.~\ref{fig:density} shows the total intensity $\ion{H}{I}$ map of our new data laid over the optical image of the galaxy. Most of the gas is settled within the central $100$ arcsec ($R\sim 29$ kpc). In the inner regions, the density of the disk is lower, producing a central depression. From the total intensity map, we infer a mass of the disk of  $M_{\rm{\ion{H}{I}}}=1.1\times 10^{10}$ M$_\odot$. These results are in agreement with the observations of \citealt{veron}. Our higher sensitivity and velocity resolution data allow us to determine the extent of the disk and to model the kinematics of the $\ion{H}{I}$. 

\begin{figure}
   \centering
   \includegraphics[width=0.49\textwidth]{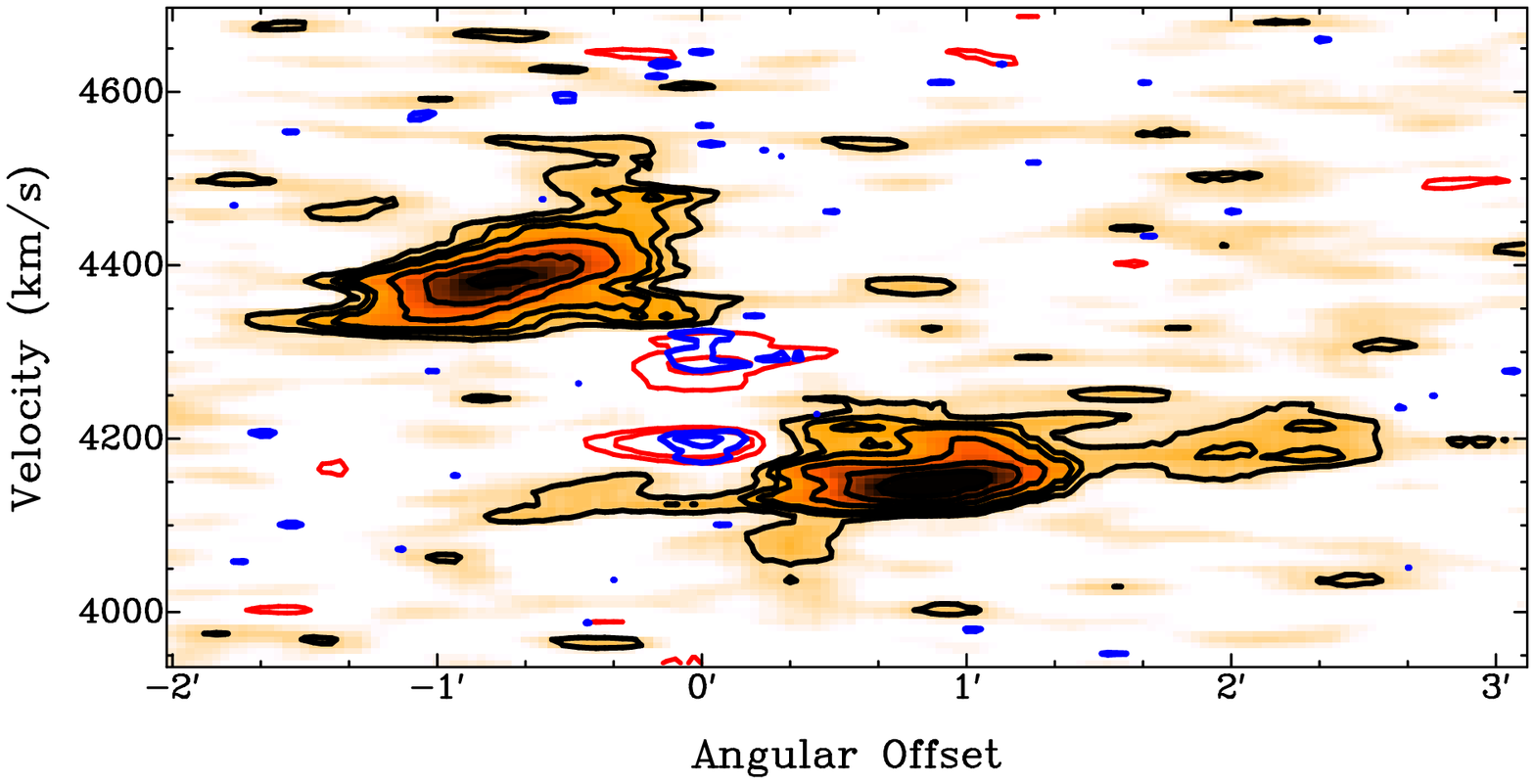}
   \caption{Position-velocity plot along a slice taken at PA$=108^\circ$ of the~\HI~disk. In \emph{black} is shown the emission. The contour levels range between $1.7$~\mJybeam and $7$~\mJybeam , with steps of $1.4$~\mJybeam. The absorption features are circled in \emph{blue} and \emph{red}, at the resolutions of $11''$ and $30''$. Contour levels are $-2,-8$~\mJybeam~and $-1.7,-3.5$~\mJybeam, respectively. The blue-shifted emission shows an external `plume' of the disk (in the NW of the total intensity map), at lower densities with respect to the regularly rotating ring. }
     \label{fig:pv} 
   \end{figure}

The position-velocity plot (Fig.~\ref{fig:pv}), shows a slice at the position angle of PA$=108^\circ$ to better highlight the kinematics of the external regions of the disk. The flatness of the rotation curve hints that, overall, the disk is regularly rotating. The blue-shifted part has a low-density feature extending up to angular offset of 2'. This corresponds to the `plume' visible in the total intensity map, N-W of the disk. Similar irregularities are also present in the southern regions of the disk. Both asymmetries hint to an ongoing merger or interaction event. These features are located in the disk at a radius \textgreater23 kpc, suggesting that these events are not directly related to the beginning of the radio activity.

The timescale of the merger event will be estimated, in the next Section, from the dynamical time of the rotating disk. 
%trim = 80 0 120 0, clip,
   %trim = 80 0 120 0, clip,

%\begin{itemize}
%\item{Other objects in the field}
%\end{itemize}
Inside the primary beam of the observations, we detected the $\ion{H}{I}$ emission of also another galaxy: ESO 102-G2\footnote{Ra$=17^{\rm{h}}21^{\rm{m}}38^{\rm{s}}$, Dec$=-65^\circ10^{\rm{'}}27^{\rm{''}}$, z$=0.0147$}. This galaxy lies $16.2$ arcmin ($285$ kpc) SW of \mbox{PKS B1718--649} at a systemic velocity of $= 4415\pm15$ km s$^{-1}$. The total inferred $\ion{H}{I}$ mass is: $M_{\rm{\ion{H}{I}}}=1.1\times10^{10}$~\msun~(\citealt{veron}) and the disk has a size of approximately $110$ arcsec. The small difference in redshift, as well as the external asymmetries of the $\ion{H}{I}$ disk of \mbox{PKS B1718--649}, slightly oriented towards ESO 102-G2, suggest a past interaction between the two galaxies. This event may have strongly influenced the kinematics of the neutral hydrogen in both galaxies.

        \begin{figure}
   \centering
   \includegraphics[width=0.5\textwidth]{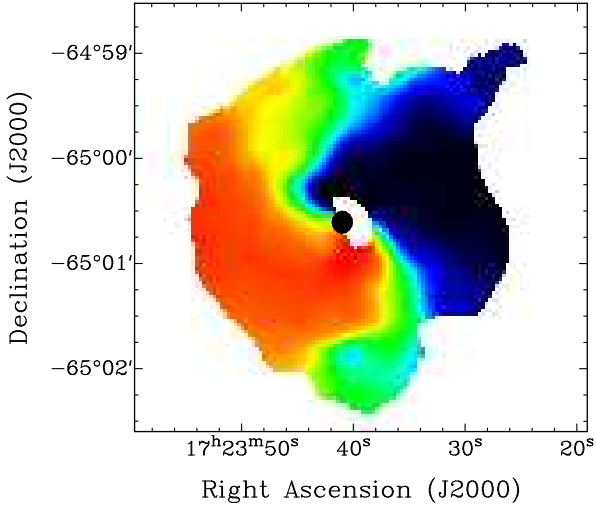}
   \caption{Velocity field of the \HI\ disk of  \mbox{PKS B1718--649}. The position angle and inclination of the minor axis of the field varies with the radius, suggesting a warped structure of the disk in the inner regions. The color-scale ranges between $4140$ \kms and $4420$\kms, showing the systemic velocity of the galaxy ($4274$ \kms) in \emph{green}. The unresolved continuum radio source is marked in \emph{black}.}
     \label{fig:vel} 
   \end{figure}  

\subsection{$\ion{H}{I}$ absorption}
\label{sec:absorption}
\begin{figure}
   \centering
   \includegraphics[width=0.5\textwidth]{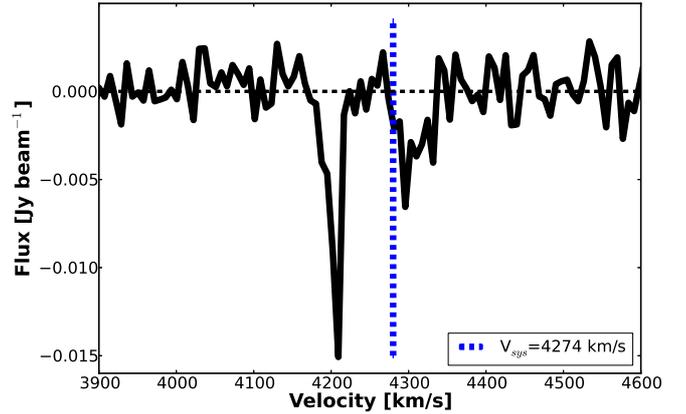}
   \caption{$\ion{H}{I}$ profile of \mbox{PKS B1718--649} obtained from the ATCA data. The two absorption systems are clearly visible. The narrow line is located at velocity $4200$ km s$^{-1}$, blue-shifted with respect to the systemic velocity ($4274$ km s$^{-1}$, \emph{blue dashed} line). The broad component is double peaked and found at velocity $4300$ km s$^{-1}$.}
     \label{fig:abs}
   \end{figure}

The $\ion{H}{I}$ absorption profile is shown in Fig.~\ref{fig:abs} and the inferred properties of the gas are summarized in Table~\ref{tab:absorption}. Two $\ion{H}{I}$ absorbing systems are detected: one is a narrow and deep (FWZI$= 43$ km s$^{-1}$, S$_{\rm{abs}}=-35.4$ mJy beam$^{-1}$), while the other is broader and shallower (FWZI$= 65$ km s$^{-1}$, S$_{\rm{abs}}=-26.0$ mJy beam$^{-1}$). The detection limit in absorption of our observations is very low (see Table~\ref{tab:datareduction}); hence, we can exclude the presence of other undetected shallow \HI~absorbing components. We determine the optical depth ($\tau$) from the following equation:

\begin{equation}
e^{-\tau}=1-\frac{S^{\rm{peak}}_{\rm{abs}}}{S_{\rm{cont}}\cdot f}
\end{equation}

\noindent where $f$ is the covering factor, assumed equal to $1$. From the optical depth it is possible to determine the column density of the inferred gas:

\begin{equation}
N_{\ion{H}{I}}=1.82\times10^{18}\cdot T_{\rm{spin}}\int \tau(v)\rm{d}v
\end{equation}

\noindent where $T_{\rm{spin}}=100$ K is the assumed spin temperature. The derived column density is $\sim7\times10^{19}$ cm$^{-2}$ for both components. The broad absorption line is red-shifted with respect to the systemic velocity (blue dashed line in Fig.~\ref{fig:abs}). On the other hand, the narrow component has a significant blue shift ($\Delta v\sim70$ km s$^{-1}$), with respect to the systemic velocity. Further insights on the nature of both absorption features will be given in Section~\ref{sec:abs}. 
%If the gas producing the $\ion{H}{I}$ absorption is located in the very inner regions of the disk, than the assumption of $T_{\rm{spin}}\sim100$ K may be unrealistic. In the case of PKS 1549-79 (Holt et al. 2006) the spin temperature is one order of magnitude higher.

\begin{table}[tbh]
\caption{Main properties of the two $\ion{H}{I}$ absorption lines.}
\centering
\begin{tabular}{l*{2}{l}} 
\hline\hline
Line & Narrow & Broad \\
\hline
$S_{\rm{peak}}$(mJy beam$^{-1}$) & --14.9 & --7.1 \\
$S_{\rm{abs}}$ (mJy beam$^{-1}$) &  --35.4 & --26.0 \\
$\tau_{\rm peak}$ & 0.004 &  0.002 \\
$\Delta v$ (km s$^{-1}$) & --74 & +26\\
$N_\ion{H}{I}$ (cm$^{-2}$) & $7.03\times10^{19} $ &  $7.74\times10^{19}$ \\
FWZI (km s$^{-1}$) & 43 & 65 \\ 
\hline
\end{tabular}\\
\label{tab:absorption}
\end{table}

\section{The modeling of the $\ion{H}{I}$ disk and the timescales of the merger}
\label{sec:discussion}
To study the global kinematics of the $\ion{H}{I}$, we fit the observed $\ion{H}{I}$ disk with a model of a regularly rotating disk, and we identify the presence of deviating components. The radius where the gas starts deviating from circular motions can be connected to the date of the last merger, or interaction event, of the disk.

The kinematics of the $\ion{H}{I}$ disks in galaxies are usually derived by fitting a tilted ring model to the observed velocity field (\citealt{rogstad}). The disk is decomposed using a set of concentric rings. Along the different radii ($r$) of these rings, the velocity of the gas is defined by three velocity components as follows:

\begin{equation}
\label{eq:vrot}
v_{\rm{obs}}(r)=v_{\rm{sys}}+v_{\rm{rot}}(r)\cos\theta\sin i+v_{\rm{exp}}\sin\theta\sin i
\end{equation}

\noindent $v_{\rm{sys}}$ is the systemic velocity at the center of each ring. $v_{\rm{rot}}(r)$ defines the rotational velocity at each radius $r$. The non-circular motions of the gas, if present, are described by $v_{\rm{exp}}$. $\theta$ denotes the azimuthal angle in the plane of the galaxy. $i$ is the inclination of each ring with respect to the line of sight. 

Different processing software, such as the Groningen Imaging Processing System ($\tt{GIPSY}$, \citealt{vanderhulst_gipsy}), provide specific routines for the tilted-ring modeling. In this paper, we use the $\tt{GIPSY}$ routine \emph{galmod} to create a model data cube, from the parameters of the rings mentioned above.

To properly estimate the physical properties of the $\ion{H}{I}$ disk, it is crucial to spatially resolve the disk with enough resolution elements. Our observations resolve the $\ion{H}{I}$ disk of \mbox{PKS B1718--649}  with only three beams on each side of the disk. This is insufficient to constrain all the parameters of Eq.~\ref{eq:vrot} with an automatic fit. Hence, we built a very schematic model using other properties of the galaxy to set constraints on the parameters. The systemic velocity is estimated combining the information given by~\citealt{veron}, the position-velocity diagram (Fig.~\ref{fig:pv}) and the global $\ion{H}{I}$ profile: $v_{\rm{sys}}=4274\pm7$ km s$^{-1}$. 

To first order, the rotational velocity is well predicted by the Tully-Fisher relation (\citealt{tully}). Given the relation in the K-band (\citealt{marc}), since the absolute magnitude of \mbox{PKS B1718--649} is $M_{\rm K20}=-24.4$, we estimate the rotational velocity of its $\ion{H}{I}$ to be $v_{\rm{rot}}=220$ km s$^{-1}$. 

We constrain the geometrical parameters of the rings from the following considerations: we center all rings at the position of the radio source (RA$=17\rm{h}23\rm{h}41.0\rm{s}$, DEC$=-65\rm{d}00\rm{m}36.6\rm{s}$). The distortion of the minor axis in the velocity field (see Fig.~\ref{fig:vel})) 
%(in \emph{green} colors in Fig.~\ref{fig:vel}) 
hints that the disk has a warped structure in the inner regions, oriented as the \halpha\ region and the dust lane (see Sec.~\ref{sec:galaxy}). A warped disk would also reproduce the steep rotation curve seen in Fig.~\ref{fig:pv}. Hence, we model a warped disk: within $r\lesssim50$ arcsec, the rings slowly switch from being N-S oriented (PA$=180^\circ$) and edge-on ($i=90^\circ$) to a more face-on and circular structure (PA$=110^\circ$, $i=30^\circ$).

From these constraints, we build the first model, which is then smoothed to the resolution of the observations. Comparing the smoothed model to the original data cube, we fine tune by hand the position angle, inclination and extension of the warped inner structure to find the best match. The complete list of the final fit parameters is shown in Table~\ref{tab:modelpar}. The quality of the fit is limited by the resolution and sensitivity of the observations. We estimate the error on the parameters to be $\pm10^\circ$ in PA and $i$ and $\pm15''$ in the extension of the warp. This includes the fact that these parameters are correlated.
%Higher spatial resolution would be needed to investigate the kinematics of the inner regions ($\lesssim 8$ kpc) in detail. 
Nevertheless, the modeling allows us to study the overall kinematics of the disk and determine the timescale of its formation.

\begin{figure}
   \centering
  \includegraphics[width=0.49\textwidth]{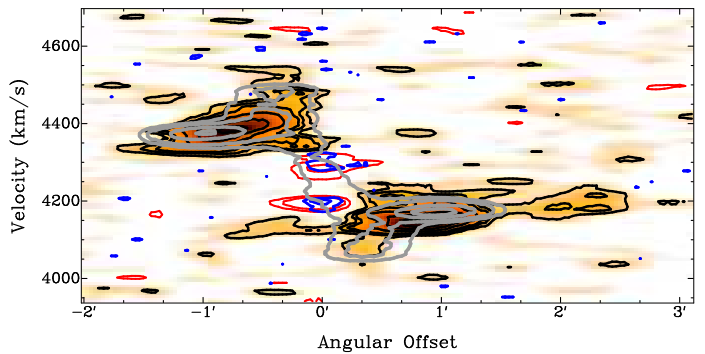}
   \caption{Position-velocity plot of the $\ion{H}{I}$ emission disk (\emph{black}) compared to the tilted ring model, smoothed at the resolution of the observations (\emph{grey} contours). The contour levels range between $1.7$~\mJybeam and $7$~\mJybeam , with steps of $1.4$~\mJybeam.  The model describes a regularly rotating disk, with a warp in the inner part, oriented along the N-S axis. The two absorption features are shown in \emph{blue} and \emph{red}, at the resolutions of $11''$ and $30''$. Contour levels are $-2,-8$~\mJybeam~and $-1.7,-3.5$~\mJybeam, respectively.}
     \label{fig:model}
   \end{figure}

The position-velocity diagram in Fig.~\ref{fig:model}, shows the same slice of Fig.~\ref{fig:pv} overlaid with the model disk (\emph{grey} contours, while the observed emission is in \emph{black}). The model disk overlaps most of the observed emission up to $R=80''$ ($R=23.5$ kpc). Therefore, our observations well match with a warped disk of neutral hydrogen settled in regular rotation up to the distance of $R=80''$, which we define as the maximum radius of regular rotation. The model does not match the observations in the outer regions: on the blue-shifted part of the rotation curve we detect emission beyond $R=80''$, in the same range of velocities of the regularly rotating component, as if it was its elongation on one side. This feature corresponds to the asymmetric `plume' mentioned in see Sec.~\ref{sec:emis}. The southern edge of the disk is also characterized by slight deviations from the regular rotation. 

The time for the \HI\ to settle into regular orbits can be assumed as the time taken by the gas to complete two revolutions around the center of the galaxy~\citep{struve}. Knowing the rotational velocity of the disk ($v_{\rm{rot}}=220$ km s$^{-1}$) and the radius of maximum regular rotation ($R=23.5$ kpc), this corresponds to  $\sim 1\times10^9$ yr.

Thus, if the \HI\ has originated via a merger or interaction event, this has occurred at least 1 Gyr ago, i.e. on a much longer timescale than the triggering of \mbox{PKS B1718--649} ($10^2$ years ago). Hence, a direct connection between these two events is unlikely.

The good agreement between the model and the observations allows us to stress that, {\sl in emission}, we do not detect, in the inner regions of \mbox{PKS B1718--649}, large clouds with significant deviations from regular rotation, and that there are no streams or radial motions in the disk which are currently bringing the cold gas close to the radio activity. Such structures may still be present in the galaxy, but they must have a mass $M_{\ion{H}{I}}<4\times10^7$ M$_\odot$, otherwise they would have been detected (see Table~\ref{tab:atca_obs}). However, the comparison with the model (Fig.~\ref{fig:model}) highlights that the two absorption lines detected against the compact radio source do not correspond to gas not regularly rotating within the disk.

%The presence of clouds with radial motions within the inner regions of the disk may also be suggested by the low-column density ($N_{\rm{\ion{H}{I}}}\sim7\times 10^{19}$ cm$^{-2}$) gas extending at anomalous blue-shifted velocities in the Figure. These clouds are not significant to modify the overall regular rotation of the disk.

\begin{table}[tbh]
\centering
\begin{tabular}{c*{4}{c}} 
\hline\hline
$r_{\rm{p}}$ & $v_{\rm{rot}}$& $i$& PA &\\
(arcsec) &  (km s$^{-1}$) &$(^\circ$))\\
\hline
5  & 220 &90 & 180 \\
10 & 220 & 90 & 180 \\
15 & 220 & 90 & 180\\
20 & 220 & 90 & 170 \\
25 & 220 & 90 & 170\\
30 & 220 & 90 & 170 \\ 
35 & 220 & 60 & 135 \\
40 & 220 & 40 & 127\\
45 & 220 & 30 & 118 \\
50 & 220 & 30 & 118 \\
55 & 220 & 30 & 118\\
60 & 220 & 30 & 118 \\
65 & 220 & 30 & 118\\
70 & 220 & 30 & 105\\
75 & 220 & 30 & 105\\
80 & 220 & 30 & 105\\
\hline
\end{tabular}\\
\renewcommand{\tablename}{Tab.}
\caption{Radially dependent parameters of the tilted ring model. $r_{p}$ represents the radius of the rings in arcsec. $v_{\rm{rot}}$ is the rotation velocity in km s$^{-1}$. $i$ and PA are the inclination and the position angle respectively, expressed in degrees. This model smoothed at the resolution of $29.7''\times27.9''$, is the best-fit to our observations.}
\label{tab:modelpar}
\end{table}

\section{The $\ion{H}{I}$ absorption and the origin of the atomic neutral hydrogen}
\label{sec:abs}
Neither of the two absorption lines is detected at the systemic velocity of the disk. Hence, since the continuum source is compact ($R<2$ pc), they cannot originate from gas regularly rotating in circular orbits. To first order, a warped disk with circular orbits describes the overall kinematics of the~\HI~disk. However, given the quality of the data, we cannot completely exclude the presence of elliptical orbits. In this scenario, the gas detected in absorption would be part of the globally rotating structure of the disk. The shifts of the two lines could be explained by gas in elliptical orbits. The presence of a barred structure in the inner regions, which may explain such orbits, is suggested by the optical morphological classification of this galaxy (SABab, see Table~\ref{tab:parameters}). It must be pointed out, though, that in emission, we are sensitive to the column densities of the observed absorption lines ($N_{\rm{\ion{H}{I}}}\sim7\times10^{19}$ cm$^{-2}$). Hence, if the absorbed gas belonged to a diffuse structure of the disk, we should have detected it also in emission.

%The combined analysis of the \HI~detected in emission and in absorption allows us to draw some hypothesis upon the nature of these clouds. 

%The disk is mostly face-on in the outer regions, hence the $\ion{H}{I}$ detected in absorption may be in proximity of the inner regions of the galaxy 

The most likely possibility is that both absorption components come from two small clouds not regularly rotating. Although these particular clouds may not have been directly involved in the triggering mechanism of the AGN, they may belong to a larger population present within the galaxy, which may contribute to fuel the AGN. These clouds could be brought in by the merger or formed for example. by cooling of the hot halo. Simulations of the accretion into the AGN, which consider the turbulence of the gas, its heating and cooling, show that cold clouds and filaments may be forming from condensation of the hot halo (\citealt{gaspari}). These clouds would chaotically collide with the surrounding medium, loosing angular momentum, falling toward the center of the galaxy and triggering the AGN.
%The past interaction event cannot also be excluded to have formed these small clouds.}

Cold gas clouds, similar to the ones of \mbox{PKS B1718--649}, have been detected also in other radio sources, which corroborates the hypothesis on their involvement in the accretion mechanisms. For example, in 3C 236 (\citealt{struve2012}) several discrete absorbing clouds, not belonging to the rotating \HI~disk, are observed close to the radio core ($\lesssim10^2$ pc). Their incoherent structure (kinetic and morphological) suggests they may be involved in the accretion mechanism into the AGN. In \mbox{NGC 315} (\citealt{morganti2009}), two different absorbing \HI~components are seen against the radio jet. One of them is associated to a cloud in-falling into the radio source. Galaxies showing ongoing interaction between the \HI~and the radio activity (e.g. via a fast cold neutral hydrogen outflow), often also show a clumpy structure of the Inter-Stellar Medium, which is populated by multiple clouds of \HI~with anomalous kinematics, (see, for example, 4C 12.50 \citealt{morganti2013,morganti2004}; and 3C 293 \citealt{mahony}). Considering that the HI disk in \mbox{PKS B1718-649} is mostly face-on in the outer regions, we also suggest that the~\HI~detected in absorption may instead belong to the inner structure of the galaxy, which is oriented edge-on.

The optical classification of radio sources between low-excitation and high-excitation radio galaxies, based on the relative strength of their optical emission lines, reflects the intrinsic difference in the efficiency of the accretion into the SMBH (\citealt{best}). HERG are radiatively efficient while LERG are radiatively inefficient accretors. The different efficiency is linked to physically different triggering mechanisms: LERG may be fueled by the cooling of the hot X-ray emitting halo of the host galaxy, while high-excitation sources may accrete cold gas, rapidly driven into the black hole by a merger or an interaction event (\citealt{hardcastle}). 
Considering  the intensity ratios of its ionized optical emission lines \citep{filippenko}, we classify \mbox{PKS B1718--649} as a low-excitation radio galaxy. Furthermore, for compact sources, it is possible to distinguish between the two different kind of AGN by measuring the ratio between the X-ray and radio luminosity of the source (\citealt{kunert}). The optical, radio and X-ray properties of \mbox{PKSB 1718-649} all suggest that \mbox{PKSB 1718-649} is a LERG. \cite{best} introduce the Eddington scaled accretion parameter $\lambda$ to estimate the accretion efficiency of LERG and HERG, which is defined as follows: 

 \begin{equation}
 \lambda=\frac{L_{\rm{mech}}+L_{\rm{rad}}}{L_{\rm{edd}}}
\end{equation}

\noindent where $L_{\rm{mech}}$ is the jet mechanical power, which is related to the $1.4$ GHz radio continuum  (\citealt{cavagnolo}), and $L_{\rm{rad}}$ is the radiative power as estimated from the [OIII] oxygen luminosity (\citealt{heckman2}). The Eddington luminosity is defined as L$_{\rm Edd}=1.3\times10^{38} M_{\rm BH}/$M$_\odot$ \ergs \citep{best}. In the case of \mbox{PKS B1718--649}, knowing that $F_{\rm\OIII}=5.0\times10^{-14}$\ergscm \citep{filippenko} and $M_{\rm BH}=4.1\times10^{8}$ \msun \citep{willett}, we determine $L_{\rm rad}=8\times10^{43}$\ergs~and $L_{\rm mech}=1\times10^{44}$\ergs. Hence, $\lambda\approx0.003$, even though the measurement is affected by the scatter in the relation between the 1.4 GHz radio power and the mechanical luminosity. The value of $\lambda$ we measured is in the range of values measured for the LERG population in the work by \cite{best}. This classification of the source would be consistent with the idea that a merger or an interaction event did not trigger this radio source.

We can investigate whether the radio nuclear activity \mbox{PKS B1718--649} could be sustained by the infall of the \HI~clouds we detected in absorption. Given the uncertainty on the origin of the clouds, we can use different approaches. The most common analytical solution of a radiatively inefficient accretion is the Bondi mechanism \citep{bondi}. Even though it makes general, unrealistic assumptions on the conditions of the accreting gas, this solution allows us to roughly estimate an upper limit to the accretion rate into the AGN, simply from the radio properties of the source. Following the empirical relation between the radio jet power and the Bondi power $\log P_{\rm j} =(1.10\pm0.11)\times\log P_{\rm B} -1.91 \pm 0.20$, measured by \cite{balmaverde}, we can estimate the mass accretion rate ($P_{\rm{B}}=\dot{M}_{\rm{Bondi}}c^2$). In the case of \mbox{PKS B1718--649}, the radio jets are undetected in the VLBI observations (\citealt{tingay2}). The $3\sigma$ noise level of the observations sets an upper limit on the flux density to $20$ mJy. Hence, the radio jet power is $P_{\rm j}\lesssim2.3\times10^{43}$\ergs~ and the Bondi power is $P_{\rm B}\lesssim1.2\times10^{45}$\ergs. Therefore we set an upper limit to the accretion rate of \mbox{PKS B1718--649} to $\lesssim0.02$~\msunyr.\\
\indent We now estimate the accretion rate that, under reasonable assumptions, we could expect to be provided by the \HI\ clouds we detect in absorption. It is plausible to assume that they have similar properties to the population of \HI~clouds detected in \mbox{NGC 315} and in \mbox{3C 236}. They might have similar sizes, $r\lesssim100$ pc, and they might be located in the innermost regions of the galaxy $\lesssim500$ pc. Under these assumptions, we can estimate a lower limit on the inflow rate of our clouds into the nuclear regions of the galaxies. Constraining their size, we determine the mass of the clouds to be $M_{\rm{\ion{H}{I}}}\gtrsim10^4$ M$_\odot$. The infall speed of the clouds into the radio core ($v_{in}\lesssim70$ km/s) is estimated from the shift of the absorption lines with respect to the systemic velocity ($\delta v$, see Table~\ref{tab:absorption}). If these clouds were located in the innermost $500$ pc, the accretion rate would be $\sim 10^{-2}$~\msunyr. Hence, a first order approximation on the accretion rate seems to suggest that the \HI~clouds detected in absorption might be able to sustain the radio activity of \mbox{PKS B1718--649}.

If the clouds we detect in \mbox{PKS B1718--649} were close to the radio source and involved in its accretion, the \HI~could have much higher temperatures than T$_{\rm{spin}}=100$ K, (see for example the case of \mbox{PKS 1549--79},~\citealt{holt}). Assuming, T$_{\rm{spin}}\sim1000$ K the column density of the absorption features would be $\sim7\times10^{20}$ cm$^{-2}$ (similarly, the minimum mass of the inferred cloud would be: M$_{\rm{\ion{H}{I}}}\gtrsim10^5 $ M$_\odot$). This agrees with the column density of the nuclear source measured from the X-ray spectrum: N$_{\rm{\ion{H}{I}}}\sim8\times 10^{20}$ cm$^{-2}$ (Siemiginowska, private communication), thus the close proximity of the \HI~clouds to the radio source may not be excluded.

%The radio core luminosity of the galaxy is $L_{\rm{core}}\sim1.4\times10^31$ W/Hz (\citealt{tingay}), which can be converted into radio power (\citealt{}): $P_{\rm{jet}}\sim8.4\times10^{37}$ W. This corresponds to an accretion power of $P_{\rm{accretion}}\sim4.6\times10^{39}$ W  (\citealt{balmaverde}). Using $P_{\rm{accretion}}=0.1\dot{M}c^2$, we determine a mass accretion rate of $\dot{M}\sim10^{-3}$ M$_\odot$.  

\section{Summary and conclusions}
\label{sec:conclusion}
We have presented new neutral hydrogen ATCA observations of the nearby young radio source \mbox{PKS B1718--649}. We detected a large $\ion{H}{I}$ disk, ($M_{\rm{\ion{H}{I}}}=1.1\times10^{10}$ M$_\odot$) with radius $R\sim29$ kpc. The disk is warped in the inner regions and overall regularly rotating. In emission, we do not detect significant deviations from the regular rotation. There are no streams or radial motions in the disk which are currently bringing the cold gas close to the radio activity. At the edges of the disk, we detect slight asymmetries, traces of a past merger or interaction event. This may have contributed in forming the settled disk, but is not related to the triggering of the radio activity. From the dynamical time of the $\ion{H}{I}$ disk, we dated the interaction event to more than $1\times10^{9}$ years ago. Since the radio source is very young ($10^2$ years), there is a significant time delay between the episode that formed the disk and the beginning of the radio activity. Even though the past interaction provided the galaxy of a massive reservoir of cold gas, the triggering of the radio activity must be attributed to another phenomenon, which let some amount of cold gas loose angular momentum and fall into the SMBH, without perturbing the overall regular rotation of the disk.

In absorption, we detect two separate lines. Their kinetic properties, compared to the ones of the~\HI~seen in emission, suggest that they may trace distinct clouds not regularly rotating within the disk. The sensitivity and resolution of the data does not allow us to determine their position, and different interpretations on their nature are possible. The link between these clouds and the radio activity cannot be excluded. Different detailed studies of accretion mechanisms (\citealt{gaspari2012,gaspari,hillel2013}) predict that accretion into the black hole can occur through chaotic collisions of small clouds of cold gas. There are also other radio sources where several clouds of cold gas, similar to the ones of \mbox{PKS B1718--649}, are detected, as, for example, \mbox{NGC 315} (\citealt{morganti2009}) and \mbox{3C 236} (\citealt{struve}). Galaxies B2 0258+35 (\citealt{shulevski}) and Centaurus A (\citealt{struve}), like \mbox{PKS B1718--649}, show a time delay between the formation of the $\ion{H}{I}$ disk and the triggering of the radio activity, and they may also have a complex \HI~structure below the overall regularly rotating disk. All these radio sources present an on-going radiatively inefficient mode of accretion. The neutral hydrogen in the form of small cold clouds, as the ones detected in \mbox{PKS B1718--649}, may play a crucial role in the triggering and fueling of this kind of radio sources.

\begin{acknowledgements}
      The Australia Telescope Compact Array is part of the Australia
Telescope which is funded by the Commonwealth of Australia for
operation as a National Facility managed by CSIRO. 
RM and FMM gratefully acknowledge support from the European Research Council under the European Union's Seventh Framework Programme (FP/2007-2013) /ERC Advanced Grant RADIOLIFE-320745.
\end{acknowledgements}

%\begin{table}[H]
%\caption{Example table}
%\centering
%\begin{tabular}{llr}
%\toprule
%\multicolumn{2}{c}{Name} \\
%\cmidrule(r){1-2}
%First name & Last Name & Grade \\
%\midrule
%John & Doe & $7.5$ \\
%Richard & Miles & $2$ \\
%\bottomrule
%\end{tabular}
%\end{table}

%------------------------------------------------

%\section{Discussion}

%\lipsum[7-8] % Dummy text

\bibliographystyle{bibtex/aa}
\bibliography{bibliorep}

\end{document}